\documentclass[twocolumn,prl]{revtex4}
\usepackage{graphicx}    
\usepackage{dcolumn}     
\usepackage{bm}          
\newcommand{\Vec}[1]{\mbox{\boldmath$#1$}}
\begin{document}
\title{Extended $s$-wave pairing originating from 
the $a_{1g}$ band in Na$_x$CoO$_2\cdot y$ H$_2$O}
\author{
Kazuhiko Kuroki$^1$, Seiichiro Onari$^2$,  
 Yukio Tanaka$^2$, Ryotaro Arita$^{3,4}$, 
Takumi Nojima$^1$
}
\affiliation{
$^1$ Department of Applied Physics and
Chemistry, The University of Electro-Communications,
Chofu, Tokyo 182-8585, Japan\\
$^2$ Department of Applied Physics, Nagoya University, 
Nagoya, 464-8603, Japan\\
$^3$ Max Planck Institute for Solid State Research, 
Heisenbergstr. 1, Stuttgart 70569, Germany\\
$^4$ Department of Physics, 
University of Tokyo, Hongo 7-3-1, Tokyo 113-0033, Japan}
\date{\today}
\begin{abstract}
Motivated by recent experiments and band calculation results, 
we investigate the possibility of superconductivity originating from 
the $a_{1g}$ band of Na$_x$CoO$_2\cdot y$H$_2$O assuming the absence of $e_g'$ 
pockets. We adopt a single band $U$-$V$ model 
that takes into account the local minimum 
of the $a_{1g}$ band dispersion at the $\Gamma$ point. Using the fluctuation 
exchange method, we find that an $s$-wave pairing with a gap that 
changes sign between the inner and the outer Fermi surfaces  
is enhanced due to 
the coexistence of spin fluctuations at wave vectors bridging the 
two Fermi surfaces and the charge fluctuations near the K point. 
Possible relevance to the experimental results is discussed.
\end{abstract}
\pacs{PACS numbers: 74.20.Mn, 71.10.Fd, 74.20.Rp, 74.25.Ha}
\maketitle

The pairing symmetry and the mechanism of 
superconductivity(SC)\cite{Takada} in a cobaltate Na$_x$CoO$_2\cdot y$H$_2$O
has attracted much attention lately.
There have been debates mainly on three issues concerning this material:
whether the relevant band is $a_{1g}$ or $e_g'$, 
whether the spin fluctuations are ferromagnetic or antiferromagnetic, 
and whether the pairing occurs in the spin-singlet 
or in the triplet channel.
In the previous studies, we have proposed 
a possibility of ferromagnetic-fluctuation mediated 
triplet $f$-wave pairing originating from the
$e'_{g}$ pocket Fermi surfaces (FS),\cite{KA,KAT1} 
whose presence was first predicted in the band calculation by 
Singh,\cite{Singh} 
and has been an issue of considerable debate in the 
band calculation studies that have followed.
\cite{JS,Pickett,Kotliar,Louie,Lee,Liebsch} 
There have also been some other theories where 
the $e_g'$ bands play an important role in the occurrence of SC.
\cite{JS2,MYO,Ikeda}

However, some recent experimental results seem to 
contradict with the $e_g'$-triplet-pairing scenario.
For instance, 
in the angle resolved photoemission (ARPES) experiments, 
the $e_g'$ bands lie below the Fermi level.
\cite{Hasan,Yang,Yang2,Takeuchi}
Also, $1/T_1T$ and Knight shift measurements suggest that the spin fluctuation
is not (or at least not purely) 
ferromagnetic in the SC samples,\cite{Fujimoto,Imai,Sato,Ishida}
and a recent theoretical calculation shows that this is 
consistent with the absence of $e_g'$ pockets.\cite{Kontani}
As for the singlet-triplet debate, recent experiments find 
a decrease of Knight shift below $T_c$ for 
magnetic fields applied parallel to the planes.\cite{Sato,Ishida}
Also, the effect of the impurities on 
$T_c$ is found to be small, suggesting a more or less isotropic pairing.
\cite{Sato2,Phillips}
Thus, there is now an increasing possibility of singlet pairing, 
although triplet pairing is still not ruled out.\cite{comment2}

Under these circumstances, 
here we investigate the possibility of SC within the 
$a_{1g}$ band {\it assuming} the absence of $e_g'$ pockets.
There already exist some theories that focus on the $a_{1g}$ band
\cite{TanakaHu,Baskaran,Kumar,Wang,Ogata,TanaOgata,Khaliullin} 
but our study differs from those at least in one of the following 
senses. (i) We take into account the 
local minimum structure of the $a_{1g}$ band dispersion at the $\Gamma$ point 
found in first principles calculations for bilayer-hydrated (BLH) systems,
\cite{JS,Kotliar,Arita}
(ii) We adopt band fillings    
corresponding to the Co valence ($V_{\rm Co}$) of $\sim +3.5$ 
because experiments show 
that the actual $V_{\rm Co}$ is much less 
than the nominal value $\sim 3.65$ due to the presence of 
H$_3$O$^+$ and/or oxygen deficiencies.
\cite{Karppinen,Takada2,Milne,Barnes,Alloul}
Combined with (i), an inner FS 
appears in addition to the outer FS, which results in 
a situation where ``disconnected FS''\cite{KA2} exist.
(Note that the Fermi level in the existing band calculations for BLH systems 
corresponds to $V_{\rm Co}\sim 3.67$, so that the inner FS does not 
exist.)
(iii) The nearest neighbor (NN) repulsion $V$, which can be important 
for systems away from half-filling, is taken into account.

We apply the fluctuation exchange (FLEX) method \cite{Bickers,Esirgen} 
to a single band $U$-$V$ model to obtain the 
Green's function, and then solve the linearized Eliashberg equation.
We find that the inner FS 
near the $\Gamma$ point strongly affects the spin fluctuations,
where the spin susceptibility takes its maximum 
at a wave vector that bridges {\it the inner and the outer FS}.
Such spin fluctuations enhance SC in a manner that the 
gap changes sign between the two FS.
The simplest among such pairing is the extended $s$-wave pairing 
whose gap does not change sign within each FS.
This extended $s$-wave pairing is further 
enhanced by charge fluctuations near the K point 
induced by the NN repulsions. We conclude that 
a combination of the 
disconnected FS and the coexistence of 
spin and charge fluctuations may lead to a $T_c$ of order $0.001t_1$
($t_1$ is the NN hopping) 
for the extended $s$-wave pairing, which roughly corresponds to the 
actual $T_c$ of the cobaltate.

\begin{figure}[t]
\begin{center}
\scalebox{0.6}{
\includegraphics[width=10cm,clip]{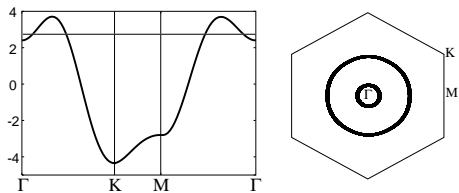}}
\caption{
Left panel: The energy dispersion for the non-interacting case with 
$t_2=-0.35$ and $t_3=-0.25$. The horizontal line shows $E_F$ for 
$n=1.5$. Right panel : FS for $n=1.5$.\label{fig1}}
\end{center}
\end{figure}

The model is given as 
${\cal H} = \sum_{i,j}\sum_{\sigma}t_{ij}c_{i\sigma}^{\dagger}c_{j\sigma}+U\sum_{i}n_{i\uparrow}n_{i\downarrow}
+\frac{1}{2}V\sum_{i,j}^{\rm NN}\sum_{\sigma\sigma'}n_{i\sigma}n_{j\sigma'}$,
where we consider hopping integrals up to third 
nearest neighbors, i.e.,  $t_1$, $t_2$, and $t_3$. 
$t_1=1$ is taken as the unit of the energy and the on-site 
repulsion is fixed at $U=6$ throughout the study.
We mainly focus on the band filling(=number of electrons/number of sites)
of $n=1.5$, which corresponds to $V_{\rm Co}=3.5$. 
In the present study,  we mainly consider the case of 
$t_2=-0.35$ and $t_3=-0.25$ (although some other values are 
also adopted for comparison), where 
the non-interacting ($U=V=0$) band dispersion has a local minimum at the 
$\Gamma$ point with a ``depth'' of O(10\%) of the total band width
\cite{comment} as seen in Fig.\ref{fig1}, so that 
an inner FS is present for $n=1.5$.

FLEX has been applied to the Hubbard model with on-site $U$,
\cite{Bickers,Dahm,Grabowski}
and has further been applied to the extended Hubbard model 
by Esirgen {\it et al.}\cite{Esirgen}.
We first obtain the renormalized Green's function, $G$, taking
bubble and ladder diagrams as the self energy.
Here, we define the ``Fermi surface'' by 
$\varepsilon_{\Vec{k}}+{\rm Re}[\Sigma_{\Vec{k}}]-\mu=0$, where 
$\varepsilon_{\Vec{k}}$is the non-interacting band dispersion, 
$\Sigma_{\Vec{k}}$ the self-energy at the lowest Matsubara 
frequency and $\mu$ the chemical potential. 
We then calculate the pairing interaction mediated by spin 
and charge fluctuations, and plug that into the linearized 
Eliashberg equation, 
$\lambda\Delta(k)=
-\frac{T}{N}\sum_{k'}\Gamma(k-k')G(k')G(-k')\Delta(k'),$
where $\Delta$ is the gap function, $G$ the Green's function, and 
$\Gamma$ the pairing interaction with $k\equiv (\Vec{k},\omega_n)$. 
$T_c$ is the temperature where the largest eigenvalue $\lambda$ reaches
unity. When the NN repulsion $V$ is introduced,  
the Green's functions and the susceptibilities become 
$7\times 7$ matrices for the triangular lattice.\cite{Esirgen}
In the actual calculation, we take up to $N=64\times 64$ sites 
and $-(2N_c-1)\pi T \leq \omega_n \leq (2N_c-1)\pi T$ 
with $N_c=2048$.

As for the pairing symmetries, we focus on $s$-wave ($A_1$ symmetry),
$d$-wave ($E_2$), and $f$-wave ($B_2$) since these 
are found to be the only pairings that have a 
chance to dominate in our model. When $V=0$ in particular, 
$f$-wave also has no chance of dominating, 
since antiferromagnetic spin 
fluctuations alone cannot mediate triplet pairing, 
so we consider only $d$- and $s$-waves.
\begin{figure}[t]
\begin{center}
\scalebox{0.6}{
\includegraphics[width=13cm,clip]{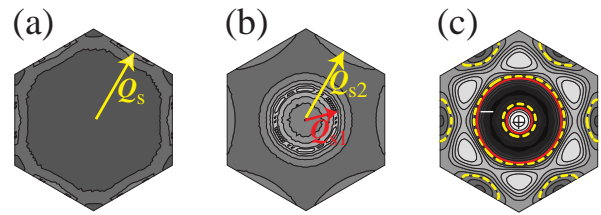}}
\caption{(color online) Contour plots of the spin susceptibility for 
(a) $t_2=t_3=0$ and (b) $t_2=-0.35$ and $t_3=-0.25$. 
(c) Contour plot the $s$-wave gap function for $t_2=-0.35$ and 
$t_3=-0.25$, where the red solid curves are the FS and 
the yellow dashed lines and curves are the nodes of the gap function.
$U=6$, $V=0$, and $T=0.01$. The gaps and the susceptibilities 
are plotted for the lowest Matsubara frequency throughout the study.
\label{fig2}}
\end{center}
\end{figure}
\begin{figure}[htb]
\begin{center}
\scalebox{0.6}{
\includegraphics[width=10cm,clip]{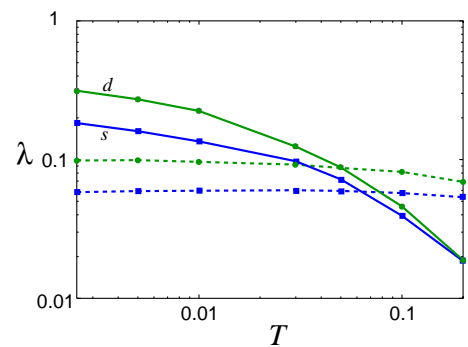}}
\caption{(color online) 
$\lambda$ plotted as functions of temperature for $s$-(blue
 lines with squares) and  $d$-waves (green lines with circles). 
The solid (dashed) lines are for $t_2=-0.35$ and $t_3=-0.25$
 ($t_2=t_3=0$). $U=6$ and $V=0$ in both cases.
\label{fig3}}
\end{center}
\end{figure}
\begin{figure}[htb]
\begin{center}
\scalebox{0.7}{
\includegraphics[width=12cm,clip]{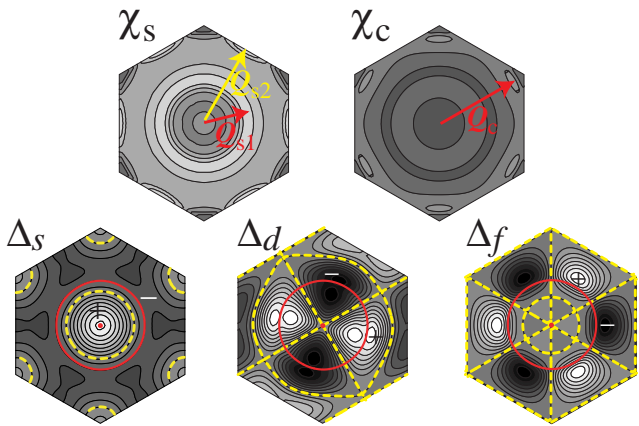}}
\caption{
(color online) Upper panel: contour plots of 
the spin $(\chi_s)$ and the charge $(\chi_c)$ susceptibilities. 
 Lower panel:
contour plots of the $s$- $(\Delta_s)$, $d$- $(\Delta_d)$, 
and $f$-wave $(\Delta_f)$ gap functions, 
where the red solid curves are the FS and 
the yellow dashed lines are the nodes of the gap function. 
$U=6$, $V=2$, $t_2=-0.35$, $t_3=-0.25$ and $T=0.025$. 
\label{fig4}}
\end{center}
\end{figure}
\begin{figure}[htb]
\begin{center}
\scalebox{0.6}{
\includegraphics[width=10cm,clip]{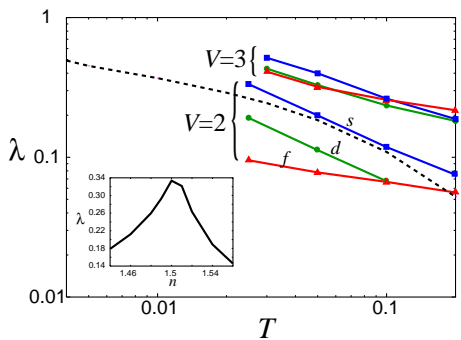}}
\caption{
(color online) $\lambda$ plotted as functions of temperature for $s$-(blue
 lines with squares), $d$- (green lines with circles), and 
$f$-wave (red lines with triangles) pairings. 
The upper (lower) lines are for $V=3$ ($V=2$). 
$U=6$, $t_2=-0.35$, and $t_3=-0.25$.  $\lambda$ for the 
$f$-wave ($B_1$ symmetry) pairing in a model for 
the $e'_g$ band is shown for comparison (dashed curve).
Inset: $n$ dependence of the $s$-wave
$\lambda$ with $V=2$ and $T=0.025$.
\label{fig5}}
\end{center}
\end{figure}
\begin{figure}[htb]
\begin{center}
\scalebox{0.6}{
\includegraphics[width=10cm,clip]{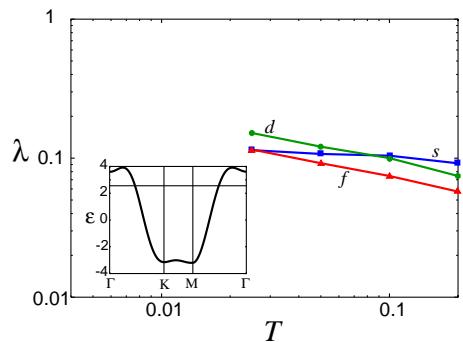}}
\caption{(color online) A plot similar to Fig.\ref{fig5} with $t_2=-0.15$ and
 $t_3=-0.25$, $V=2$. 
inset: the energy dispersion for $U=V=0$, 
$t_2=-0.15$, $t_3=-0.25$, $U=6$, $V=2$, and $T=0.025$.
\label{fig6}}
\end{center}
\end{figure}

Let us move on to the calculation results. 
We first show results without the NN 
repulsion $V$. When $t_2=t_3=0$, 
the spin susceptibility 
broadly peaks around the wave vectors that bridge the 
opposite sides of the FS as seen in Fig.\ref{fig2}(a).
Both $d$-wave and $s$-wave pairings have very small values of $\lambda$ 
(Fig.\ref{fig3}),  
suggesting that there is no SC instability for $n=1.5$ 
at least in the realistic temperature range.
When we turn on the distant hoppings $t_2=-0.35$ and $t_3=-0.25$ so that 
the inner FS appears,
the spin susceptibility still has peaks at the Brillouin zone edge 
($\Vec{Q}_{s2}$), but 
takes larger values at wave vectors $\Vec{Q}_{s1}$ that bridges the 
outer and the inner FS as seen in Fig.\ref{fig2}(b).
A repulsive pairing interaction mediated by the 
spin fluctuations around $\Vec{Q}_{s1}$ can enhance SC 
because a sign change of the gap between the inner and the 
outer FS can take place without
adding additional nodes on the FS.
In fact, $\lambda$ of both 
extended $s$-wave (gap shown in Fig.\ref{fig2}(c)) and 
$d$-wave (gap not shown) pairings are enhanced by a 
factor of 3 to 4 at low temperature as seen in Fig.\ref{fig3}. 
Nevertheless, these values are still much smaller than unity even at 
$T=0.0025$, which roughly corresponds to the actual $T_c$ of the 
cobaltate. 

This is where the NN repulsion $V$ sets in.
The FLEX results in the presence of $V$ 
show that while the spin susceptibility $\chi_s$
has structures similar to those in the absence of $V$, 
the charge susceptibility $\chi_c$ peaks near the K point (Fig.\ref{fig4}), 
which bridges the opposite sides of 
the outer FS. This is because the Fourier transform of the 
NN repulsion takes a large negative value at K point. 
Such charge fluctuations work cooperatively 
with the spin fluctuations around $\Vec{Q}_{s1}$ to 
enhance the extended $s$-wave pairing having the same gap sign 
within each FS ($\Delta_s$ in Fig.\ref{fig4}). This is because the 
charge fluctuations mediate attractive pairing interactions 
as opposed to the spin fluctuations.

In fact, the eigenvalue of the $s$-wave pairing 
is strongly enhanced in the presence of $V$
as seen in Fig.\ref{fig5}.
$d$- and $f$-wave pairings are also enhanced because these 
gaps can exploit the attractive interaction near the K point 
at some portions of the FS,\cite{TanaOgata} 
but $s$-wave pairing dominates over $d$ and $f$ at least for the parameter 
values adopted here.
Although we cannot go down to low temperatures in the presence of $V$,
the nearly linear increase of $\log(\lambda)$ with 
decreasing $\log(T)$ 
suggests that $\lambda$ may reach unity 
at least around $T=O(0.001t_1)$ for the $s$-wave pairing.
Note that the values of $\lambda$ for 
the $s$-wave pairing 
is even larger than that 
for the $f$-wave pairing ($B_1$ symmetry) state originating from the 
$e_g'$ pockets analyzed in a single band model in ref.\cite{KAT1},
at least for the temperature range where the calculation is 
performed.

The inner FS in Fig.\ref{fig4} is smaller than 
that in Fig.\ref{fig1}, which means that the correlation effect makes the 
depth of the local minimum of the band shallower,  
so that the Fermi level is located very close to the local bottom.
We find that the values of $\lambda$ are not sensitive to the choice of the 
values of the distant hoppings as far as this relation between the 
Fermi level and the energy at the $\Gamma$ point 
does not change 
significantly. For instance, if we take $t_2=-0.45$, $t_3=-0.1$,
and $t_4=-0.04$, where $t_4$ is the fourth nearest neighbor hopping, 
the relation between the Fermi level and the local minimum at $\Gamma$ 
is similar to that for 
$t_2=-0.35$ and $t_3=-0.25$, but the outer FS 
has a more hexagonal shape rather than the round 
one shown in Fig.\ref{fig1}. Performing FLEX calculation for 
this set of parameter values, we find that the values of 
$\lambda$ are very close to those shown in Fig.\ref{fig5}.

On the other hand, the situation 
changes drastically if the local minimum 
is too shallow to have an inner FS at $n=1.5$, as 
expected from the results for $V=0$ (Fig.\ref{fig3}).  
We show in Fig.\ref{fig6} the 
calculation results for $t_2=-0.15$ and $t_3=-0.25$,  
where the local minimum of the band at the $\Gamma$ point 
lies far above the Fermi level for $n=1.5$ as seen in the inset.
It can be seen that the $s$-wave state 
is strongly suppressed compared to the case of $t_2=-0.35$, 
and none of the pairing symmetries are likely to have a $T_c$ in 
the realistic temperature range.

We further find that when the local minimum is too deep (by taking, 
for instance, a too large $-t_2$), so that 
the inner FS becomes large, the $s$-wave pairing 
is again suppressed. This is because the distance between 
the two FS becomes short, so that the 
gap nodes are located too close to the FS.
Such a tendency can also be captured by fixing $t_2=-0.35$, $t_3=-0.25$, 
and varying $n$, where we find that $\lambda$ is maximized 
around $n=1.5$ (the inset of Fig.\ref{fig5}).
These results show that SC can be 
realized (in the realistic temperature range) only when the 
Fermi level is located near the local bottom of the band.

Finally, 
let us discuss the relevance of the present results to the experiments. 
(i) SC, accompanied by enhanced 
non-ferromagnetic (or incommensurate) 
spin fluctuations\cite{Fujimoto,Imai,Ishida2,Ishida}, 
is observed {\it only} in the BLH samples.\cite{Takada3} 
In the present view, this can be explained as follows. 
In the non-hydrated and MLH systems, there is a large three dimensionality 
in the $a_{1g}$ bands, as can be seen from the split of the two bands
(originating from two CoO$_2$ layers in a unit cell) at the $\Gamma$ point.
\cite{JS,Kotliar,Arita}
This results in one band without the inner FS and the other with 3D inner 
and 2D outer FS, which would smear out the $Q_{s1}$ spin fluctuations 
discussed here. Moreover, for non-hydrated systems in particular, 
the local minimum depth tends to be small,\cite{JS} so that there may be no 
inner FS at all.
(ii) $s$-wave pairing 
is consistent with the absence of time reversal symmetry breaking
suggested from $\mu$SR experiments.\cite{Higemoto,Uemura}
(iii) $s$-wave pairing is also consistent with 
the decrease of Knight shift below $T_c$.\cite{Sato,Ishida} 
(iv) Although the nodes of the gap do not intersect the 
FS in the extended $s$-wave state, the change of both the 
sign and the absolute value of the gap between the inner and the outer 
FS may explain some of the unconventional behavior observed in $1/T_1$ 
\cite{Fujimoto,Ishida3,Sato} or heat capacity measurements.
\cite{Jin,Phillips} 
In fact, we have preliminary results showing 
that the absence of a coherence peak and the crossover from $\sim T^3$ 
to $\sim T$ behavior in $1/T_1$ 
can be understood within the present $s$-wave 
pairing by taking into account the effect of impurities.
(v) The extended $s$-wave pairing may also 
account for the small impurity effect on $T_c$\cite{Sato2,Phillips} 
since $\sum_{\Vec{k}\in{\rm FS}}\Delta(\Vec{k})=0$ is not satisfied 
due to the asymmetry between the two FS.
(vi) It has recently been found that $T_c$ is maximized around
$V_{\rm Co}\sim 3.5$.\cite{Barnes} This is consistent with our conclusion 
that SC can be realized only when the Fermi level is
located near the local bottom of the band 
(inset of Fig.\ref{fig5};note that $V_{\rm Co}=5-n$).

To summarize, 
we have shown that an extended $s$-wave pairing can arise 
from the $a_{1g}$ band due to a 
combination of disconnected inner and outer FS and 
coexisting spin and charge fluctuations. The present scenario seems to be 
consistent with many of the experimental results.

We acknowledge K. Held for valuable discussions and critical reading of the 
manuscript. We thank Y. Kobayashi and 
H. Sakurai for discussions on experimental results.
K.K. also thanks Y. Hattori and M. Tanaka for discussions on the 
band dispersion. Numerical calculation has been performed
at the facilities of the Supercomputer Center,
Institute for Solid State Physics, University of Tokyo, and 
at the Information Technology Center, University of Tokyo.
This study has been supported by 
Grants-in-Aid for Scientific Research from the Ministry of Education, 
Culture, Sports, Science and Technology of Japan, and from the Japan 
Society for the Promotion of Science.

%



\begin{thebibliography}{99}
\bibitem{comment2} Whether the Knight shift decreases 
for magnetic fields perpendicular 
to the planes is controversial at the present stage. The 
magnitude of the upper critical field is also an issue of controversy.
\bibitem{comment} The depth depends on band calculations and thus has  
some ambiguity. Our standpoint here
is that in the BLH samples, this depth should be large enough 
so that the Fermi level touches the local bottom of the band 
(the necessary depth depends on the actual 
$V_{\rm Co}$, which is also somewhat ambiguous at the present stage) 
because otherwise it would be difficult to understand the 
enhancement of non-ferromagnetic spin fluctuations, which is not 
observed in non-hydrated samples.\cite{Imai,Ishida2}
\end{thebibliography}
\end{document}